\documentclass[12pt]{iopart}
\usepackage{epsfig,graphics}
\begin{document}

\title[Comparison of GT strengths in the RPA]{Comparison of Gamow-Teller strengths in the random phase approximation}

\author{Jameel-Un Nabi}
\address{Faculty of Engineering Sciences, GIK Institute of
Engineering Sciences and Technology, Topi 23640, Khyber Pakhtunkhwa,
Pakistan,\\ and \\
The Abdus Salam ICTP, Strada Costiera 11, I-34151, Trieste, Italy}

\author{Calvin W. Johnson}
\address{Department of Physics, San Diego State University,
5500 Campanile Drive, San Diego, CA 92182-1233}

\begin{abstract}
The Gamow-Teller response is astrophysically important for a number
of nuclides, particularly around iron. The random phase
approximation (RPA) is an efficient way to generate strength
distributions. In order to better understand both theoretical
systematics and uncertainties, we compare the Gamow-Teller strength
distributions for a suite of nuclides and for a suite of
interactions, including semi-realistic interactions in the $1p$-$0f$
space with the RPA and a separable multi-shell interaction in the
quasi-particle RPA. We also compare with experimental results
where available.
\end{abstract}
\pacs{21.10.Pc, 21.60.Cs,  21.60.Jz} \maketitle

\section{Introduction}
Gamow-Teller (GT) electron capture ($\beta$-decay) transitions,
caused by the $\sigma\tau_+$ ($\sigma\tau_-$) operator, are some of
the most important nuclear weak processes in astrophysics. For a
review of spin-isospin transitions see Ref. \cite{Ost92}. The GT
transitions in $fp$-shell nuclei play important roles at the core
collapse stages of supernovae, specially in neutrino induced
processes. One of the factors controlling the gravitational
core-collapse of massive stars is the lepton fraction; the lepton
fraction in turn is governed by $\beta$-decay and electron capture
rates among iron-regime nuclides. A primary and non-trivial
contribution to the weak rates is the distribution of GT strength.
GT strengths have important implications in other astrophysical
scenarios as well, such as explosive nucleosynthesis in O-Ne-Mg
white dwarfs (see Ref. \cite{NR07} and references therein) .

GT distributions have been extracted experimentally using different
techniques. Whereas the $\beta$-decay extraction is done in a model
independent manner and are used to calibrate the B(GT), the
charge-exchange reactions require further assumptions and the
resulting extraction of GT distributions cannot be truly done in a
model-independent manner. Consequently astrophysical calculations
rely either upon crude estimates or upon more detailed microscopic
calculations. The main difficulty with both experiment and theory is
that the strength distribution connects to many states. Further in
astrophysical environments one needs finite temperature GT strength
functions as the temperature is high enough for excited states in
the parent to be thermally populated.

The isovector response of nuclei may be studied using the nucleon
charge-exchange reactions $(p,n)$ or $(n,p)$; by other reactions
such as $(^{3}$He,$t$), ($d,^{2}$He) or through heavy ion reactions.
The $0^{0}$ GT cross sections ($\Delta T =1, \Delta S =1, \Delta L
=0, \hspace{0.1cm} 0\hbar\omega$ excitations) are proportional to
the analogous beta-decay strengths. Charge-exchange reactions at
small momentum transfer can therefore be used to study beta-decay
strength distributions when beta-decay is not energetically
possible. The $(p,n)$,  $(^{3}$He,$t)$ reactions probe the GT$_{-}$
strength (corresponding to $\beta^{-}$-decay) and the $(n,p)$,
$(d,^{2}$He) reactions give the strength for
$\beta^{+}$-decay/electron capture, i.e. GT$_{+}$ strength. The
study of $(p,n)$ reactions has the advantage over $\beta$-decay
measurements in that the GT$_{-}$ strength can be investigated over
a large region of excitation energy in the residual nucleus. On the
other hand the $(n,p)$ reactions populates only $T = T_{0}+1$ states
in all nuclei heavier than $^{3}$He. This means that other final
states (including the isobaric analog resonance) are forbidden and
GT$_{+}$ transitions can be observed relatively free of background.
The study of these reactions suggest that a reduction in the amount
of GT strength is observed relative to theoretical calculations. The
GT quenching is on the order of 30-40 $\%$ \cite{Vet89}.

Theory for GT transitions falls generally into three camps: simple
independent-particle models (e.g.  Ref. \cite{Ful80}); full-scale
interacting shell-model calculations; and, in between, the
random-phase approximation (RPA) and quasi-particle random-phase
approximation (QRPA). Independent-particle models underestimate the
the total GT strength, because the Fermi surface is insufficiently
fragmented, while  also placing the centroid of the GT strength too
high for  even-even parent nuclides and too low  on odd-A and
odd-odd parents \cite{Pin00}. Full interacting shell-model
calculations are computationally demanding, although one can exploit
the Lanczos algorithm, commonly used in large shell-model
diagonalization \cite{Whi72}, to efficiently generate the strength
distribution \cite{SMreview05}; for medium-mass nuclei one still
needs to choose from among a number of competing
semi-realistic/semi-empirical interactions. RPA and QRPA can be
thought of as approximations to a full shell-model calculation and
are much less demanding computationally.

In this paper we compare GT strength distributions for a suite of
iron-region nuclides relevant to astrophysics: $^{54,55,56}$Fe,  and
$^{56,58}$Ni. For each of these nuclides we compute the GT strengths
in several RPA calculations. Each RPA calculation is in occupation
rather than configuration space, which is appropriate inasmuch as
the GT operator only affects spin and isospin. (In fact each
calculation properly speaking is proton-neutron RPA or QRPA, as the
RPA/QRPA phonon operators change protons into neutrons or
vice-versa.) The calculations, which will be described in greater
detail in subsequent sections, are :

\noindent $\bullet$ pn-RPA in a major harmonic oscillator shell,
that is, the $1p$-$0f$ shell, with three different
semi-realistic/semi-empirical interactions \cite{Pov01,Ric91,Hon02}.
For details see Section 2.

\noindent $\bullet$ pn-QRPA in a multi-shell single-particle space
with a schematic interaction that has been previously applied to
similar calculations \cite{Mut92}. For further details we refer to
Section 3.

These particular calculations were chosen because of the availability of codes; one could
imagine a larger set of calculations (e.g., pn-QRPA with semi-realistic interactions)
but relevant codes either do not exist or are not available to us.

These calculations will help to understand systematic similarities
and differences between (a) different $1p$-$0f$ shell-model
interactions and (b) between $0\hbar\Omega$ shell-model calculations
against multi-shell calculations with a separable interaction. For
example, for some cases in the $1s$-$0d$ shell the separable
interactions yield a larger total strength and a higher centroid
\cite{Nab99} than shell-model calculations.

Section~4 presents the results and discussions on use of various
pn-RPA schemes. We finally present the summary and conclusions in
Section~5.

\section{The random phase approximation with shell-model interactions}

The configuration-interaction (CI)  shell model solves the many-body
problem in a large basis of Slater determinants using the occupation
representation.  One advantage of the CI shell model is that it can
use arbitrary two-body (or even higher-order) interactions and gives
explicit wavefunctions for excited states as well as the ground
state. In addition, because the GT operator is $\sigma \tau$ and
does not affect coordinate space wavefunctions, the CI shell model
is well suited for GT transitions. The drawback of the CI shell
model is that even with including just a few, or even one, harmonic
oscillator shell, the basis dimensions can be huge ($10^9$ or
greater), making such calculations computationally intensive. Furthermore, limiting
calculations to a few shells leads to the necessity for quenching (in the
case of Gamow-Teller transitions) or enhancement (for example, E2 transitions)
of coupling parameters.

The Hamiltonian for the shell-model is written in occupation space\cite{Rin80,BG77,SMreview05}, i.e.,
\begin{equation}
\hat{H} = \sum_a \epsilon_a \hat{n}_a +\frac{1}{4}\sum_{abcd}
V_{abcd} \hat{c}^\dagger_a \hat{c}^\dagger_b \hat{c}_c \hat{c}_d
\end{equation}
where the creation and annihilation operators $\hat{c}_a^\dagger, \hat{c}_a$ represent single-particle states with good angular momentum, and where $\epsilon_a$ are the single particle energies and
$V_{abcd}$ are the two-body matrix elements.

It is possible to solve the RPA matrix equation in a shell-model
representation, using the single-particle energies and two-body
matrix elements above. A recent series of papers showed that RPA is
a reasonable, if not perfect, approximation to the numerically exact
results, comparing ground state correlation energies
\cite{stetcu2002} and charge-conserving \cite{stetcu2003} and
charge-changing \cite{stetcu2004} transitions; in the last case it
was found that allowing the Hartree-Fock state to be deformed
improved pn-RPA calculations of GT strength distributions.

The first step is a Hartree-Fock calculation, which introduces a unitary transformation on the
single-particle states,
\begin{equation}
\hat{d}^\dagger_\alpha = \sum_a D_{\alpha a} \hat{c}^\dagger_a
\end{equation}
These states are divided into occupied (hole) states, labeled by
$m$, and unoccupied (particle) states labeled by $i$, and the
transformation matrix $\mathbf{D}$ is chosen such that the energy of
the Slater determinant $ | \mathrm{HF} \rangle = \prod_m
\hat{d}^\dagger_m | 0 \rangle$ minimizes the energy $\langle
\mathrm{HF} | \hat{H} | \mathrm{HF} \rangle$.

With the Hartree-Fock solution  in hand, one finds excited states
(and the correlation energy in the ground state, although that does
not concern us here) by treating the energy surface in the vicinity
of the Hartree-Fock state as quadratic.  This leads to the RPA
matrix equations \cite{Rin80}.  For charge-changing interactions
such as Gamow-Teller, the RPA matrix equations take the form:
\begin{equation}
\left(\begin{array}{cccc}
A^{np,pn} & 0 & 0 & B^{np,pn} \\
0 & A^{pn,np} & B^{pn,np} & 0 \\
0 & -B^{np,pn} & -A^{np,pn} & 0 \\
-B^{pn,np} & 0 & 0 & -A^{pn,np}
\end{array}\right)\left(\begin{array}{c}
X(pn)\\
X(np)\\
Y(np)\\
Y(pn)\end{array}\right)=\Omega\left(\begin{array}{c}
X(pn)\\
X(np)\\
Y(np)\\
Y(pn)\end{array}\right),
\label{pnRPAeq}
\end{equation}
where the definitions for $A^{pn,np}$ and $B^{np,pn}$
matrices are similar to the regular proton-neutron conserving formalism,
where one approximates $ | \mathrm{RPA} \rangle \approx
| \mathrm{HF} \rangle$:
\begin{equation}
A_{mi,nj}^{np,pn}=\langle {\rm HF}|[\nu_i^\dagger \pi_m, [H,\pi_n^\dagger\nu_j]]|
{\rm HF} \rangle=(\epsilon^p_n-\epsilon_i^n)\delta_{mn}\delta_{ij}
-V^{pn}_{mn,ji},
\label{defA}
\end{equation}
\begin{equation}
B_{mi,nj}^{np,pn}=-\langle {\rm HF}|[\nu_m^\dagger \pi_i,[\pi_n^\dagger\nu_j,H]]|
{\rm HF} \rangle=-V_{in,jm}^{pn}.
\label{defB}
\end{equation}
The matrices $A^{pn,np}$ and $B^{pn,np}$ are defined similarly,
but are distinct unless $Z=N$; in fact, they have different dimensions
unless $Z=N$. Let $N_p^\pi$,
$N_h^\pi$ be number of proton particle and hole states, respectively, and
$N_p^\nu$, $N_h^\nu$ the number of neutron particle and hole states. Thus
the vectors $X(pn)$ and $Y(np)$ are of length $N_p^\pi N_h^\nu$ while
vectors $X(np)$, $Y(pn)$ are of length $N_p^\nu N_h^\pi$; the two lengths are
unequal unless $Z=N$. Similarly,
$\mathbf{A}^{np,pn}$ is a square matrix of dimension $N_p^\pi N_h^\nu$  while
$\mathbf{A}^{pn,np}$  is a square matrix of dimension $N_p^\nu N_h^\pi$,
while $\mathbf{B}^{np,pn}$ is a rectangular matrix of dimension $N_p^\pi N_h^\nu \times N_p^\nu N_h^\pi$,
and   $\mathbf{B}^{pn,np}= \left(\mathbf{B}^{np,pn}\right)^T$ .

The transition strength is given by
\begin{eqnarray}
\langle \mathrm{RPA} | {\cal O}|\lambda_{(Z\pm 1,N\mp
1)}\rangle=\langle \mathrm{RPA} | [{\cal
O},\beta_\lambda^\dagger]|\mathrm{RPA}\rangle \nonumber \\ =
\sum_{mi}\left(X_{mi}^\lambda(pn/np) {\cal
O}_{mi}+Y_{mi}^\lambda(pn/np){\cal O}_{im}\right),
\end{eqnarray}
where $\beta_\lambda^\dagger$ is the transition operator for a $\beta$-decay,
in this case the Gamow-Teller operator $\vec{\sigma} \tau_\pm$.
For more details consult \cite{stetcu2004}.

For this paper we use three different semi-realistic/semi-empirical
shell model interactions. All three interactions started from
realistic nucleon-nucleon interactions, from which an effective
interaction (e.g., a G-matrix) was derived. At this point the
interaction is expressed numerically as two-body matrix elements
$V_{JT}(ab,cd)$.  The interactions were all then further modified in
order to fit experimental spectra; as is well-known to the
shell-model community, most of the modification were to the
``monopole'' parts of the interaction, which are related to
properties of the mean-field.  All three interactions are similar,
but have different starting points and were fitted to different data
sets, with the following semi-realistic/semi-empirical interactions:
the modified Kuo-Brown interaction KB3G \cite{Pov01} and the
Brown-Richter  interaction interaction FPD6G \cite{Ric91} and the
Tokyo interaction  GXPF1 \cite{Hon02}; the names do not signify much
except that PF/FP refer to the $fp$ shell.

\section{The quasi-particle random phase approximation with a separable interaction}

For an alternate approach, we used the quasi-particle proton-neutron
random phase approximation ($pn$-QRPA) with a separable interaction
of the form

\begin{equation}
\hat{H}^{QRPA} = \hat{H}^{sp} + \hat{V}^{pair} + \hat{V}_{GT}^{ph} + \hat{V}_{GT}^{pp}
\end{equation}
where $\hat{H}^{sp}$ is the single-particle Hamiltonian, $\hat{V}^{pair}$ is the
pairing force, $\hat{V}^{ph}_{GT}$ and $\hat{V}^{pp}_{GT}$ are the particle-hole (ph)
and particle-particle (pp) components, respectively, of the GT force $(\vec{\sigma}
\vec{\tau} )^2$.  We
diagonalized our Hamiltonian in three consecutive steps as outlined
below.

Single-particle energies and wave functions were calculated in the
Nilsson model which takes into account nuclear deformation
\cite{Nil55}. The transformation from the spherical basis to the
axial-symmetric deformed basis can be written as \cite{Mut92}
\begin{equation}
\hat{d}^{\dagger}_{m\alpha} = \sum_{j}D^{m\alpha}_{j}\hat{c}^{\dagger}_{jm},
\end{equation}

where $\hat{d}^{\dagger}$ and $\hat{c}^{\dagger}$ are particle creation operators in the
deformed and spherical basis, respectively; the transformation
matrices $D^{m\alpha}_{j}$ were determined by diagonalization of the
Nilsson Hamiltonian, and $\alpha$ represents additional quantum
numbers, except $m$, which specify the Nilsson eigenstates.

Pairing was treated in the BCS approximation, where a constant
pairing force with the force strength $G$ ($G_{p}$ and $G_{n}$ for
protons and neutrons, respectively) was applied,
\begin{equation}
\hat{V}^{pair} =
-G\sum_{jmj^{\prime}m^{\prime}}(-1)^{l+j-m}\hat{c}^{\dagger}_{jm}\hat{c}^{\dagger}_{j-m}
(-1)^{l^{\prime}+j^{\prime}-m^{\prime}}\hat{c}_{j^{\prime}-m^{\prime}}\hat{c}_{j^{\prime}m^{\prime}},
\end{equation}

where the sum over $m$ and $m^{\prime}$ was restricted to
$m,m^{\prime} > 0$, and $l$ represents the orbital angular momentum.

The BCS calculation gave the quasi-particle energies
$\epsilon_{m\alpha}$. A quasi-particle basis was introduced via
\begin{equation}
\hat{a}^{\dagger}_{m\alpha} = u_{m\alpha}\hat{d}^{\dagger}_{m\alpha} -
v_{m\alpha}\hat{d}_{\overline{m}\alpha},
\end{equation}
\begin{equation}
\hat{a}^{\dagger}_{\overline{m}\alpha} = u_{m\alpha}\hat{d}^{\dagger}_{\overline{m}\alpha} +
v_{m\alpha}\hat{d}_{m\alpha},
\end{equation}

where $\overline{m}$ is the time-reversed state of $m$, and
$\hat{a}^{\dagger} (\hat{a})$ are the quasi-particle creation
(annihilation) operators which enter the RPA equation. The
occupation amplitudes $u$ and $v$ satisfy the condition $u^{2} +
v^{2} = 1$ and were determined by the BCS equations (see for example
\cite{Rin80}, page 230).

In the pn-QRPA, charge-changing transitions are expressed in terms
of phonon creation, with the QRPA phonons defined by
\begin{equation}
\hat{b}^{\dagger}_{\omega}(\mu) =
\sum_{pn}(X^{pn}_{\omega}(\mu)\hat{a}^{\dagger}_{p}\hat{a}^{\dagger}_{\overline{n}}
- Y^{pn}_{\omega}(\mu)\hat{a}_{n}\hat{a}_{\overline{p}})
\label{phonon}.
\end{equation}

The sum in Eq.~(\ref{phonon}) runs over all proton-neutron pairs
with $\mu = m_{p} - m_{n} =$ -1, 0, 1, where $m_{p(n)}$ denotes the
third component of the angular momentum. The ground state of the
theory is defined as the vacuum with respect to the QRPA phonons,
$\hat{b}_{\omega}(\mu)|QRPA> = 0$. The forward- and backward-going
amplitudes $X$ and $Y$ are eigenfunctions of the RPA matrix equation

\begin{equation}
 \left[ \begin{array}{cc}
 A & B \\  -B & A
 \end{array} \right]  \left[ \begin{array}{c} X \\ Y \end{array} \right] =
 \omega \left[ \begin{array}{c} X \\ Y \end{array} \right], \label{RPA}
\end{equation}
where $\omega$ are energy eigenvalues of the eigenstates and
elements of the two submatrices are given by
\begin{eqnarray}
A_{pn,p^{\prime}n^{\prime}} & = & \delta(pn, p^{\prime}n^{\prime})(\epsilon_{p}+\epsilon_{n}) + V^{pp}_{pn,p^{\prime}n^{\prime}}(u_{p}u_{n}u_{p^{\prime}}u_{n^{\prime}}+v_{p}v_{n}v_{p^{\prime}}v_{n^{\prime}}) + \nonumber \\
& & V^{ph}_{pn^{\prime},
p^{\prime}n^{\prime}}(u_{p}v_{n}u_{p^{\prime}}v_{n^{\prime}}+v_{p}u_{n}v_{p^{\prime}}u_{n^{\prime}}),
\label{RPA_A}
\end{eqnarray}
\begin{equation}
\label{RPA_B}
 B_{pn,p^{\prime}n^{\prime}} =
V^{pp}_{pn,p^{\prime}n^{\prime}}(u_{p}u_{n}v_{p^{\prime}}v_{n^{\prime}}+v_{p}v_{n}u_{p^{\prime}}u_{n^{\prime}})
-V^{ph}_{pn,
p^{\prime}n^{\prime}}(u_{p}v_{n}v_{p^{\prime}}u_{n^{\prime}}+v_{p}u_{n}u_{p^{\prime}}v_{n^{\prime}}).
\end{equation}

The backward-going amplitude $Y$ accounts for the ground-state
correlations. It is essential to note however, that the derivation
of the QRPA matrix requires ground-state correlations to be only a
small correction. It should be noted that $|Y| \ll |X|$ does not
imply that ground-state correlations are negligible, since for the
calculation of $\beta$ transition matrix elements one always must
consider products of the form $uvY$ and $u^{\prime}v^{\prime}X$.
Especially in $\beta^{+}$ decay, $uv$ can be larger than
$u^{\prime}v^{\prime}$; thus ground-state correlations cannot be
neglected. The RPA equation is constructed and solved for each value
of the projection $\mu$, i.e., $\mu =$ -1, 0 and +1. The equation
gives identical eigenvalue spectra for $\mu =$ -1 and $\mu =$ +1,
and eigenvalues for $\mu =$ 0 are always two-fold degenerate,
because of the axial symmetry of the Nilsson potential. (Hereafter,
$\mu$ will be suppressed if not otherwise stated, since the
following formulas hold for each $\mu$).

In the pn-QRPA formalism proton-neutron residual interactions occur
in two different forms, namely as particle-hole (ph) and
particle-particle (pp) interaction. Both the particle-hole and
particle-particle interaction can be given a separable form.

In the present work, in addition to the well known particle-hole
force \cite{Hal67,Sta90}
\begin{equation}
 \hat{V}^{ph}_{GT} = 2\chi
\sum_{\mu}(-1)^{\mu} \hat{Y}_{\mu} \hat{Y}^{\dagger}_{-\mu},
\end{equation}
with
\begin{equation}
 \hat{Y}_{\mu} =
\sum_{j_{p}m_{p}j_{n}m_{n}}\langle
j_{p}m_{p}|t_{-}\sigma_{\mu}|j_{n}m_{n} \rangle
 \hat{c}^{\dagger}_{j_{p}m_{p}} \hat{c}_{j_{n}m_{n}},
\end{equation}
the particle-particle interaction, approximated by the separable
force \cite{Sol87, Kuz88}
\begin{equation}
 \hat{ V}^{pp}_{GT} = -2\kappa
\sum_{\mu}(-1)^{\mu} \hat{P}_{\mu}^{\dagger} \hat{P}_{-\mu},
\end{equation}
 with
\begin{equation}
 \hat{P}^{\dagger}_{\mu} =
\sum_{j_{p}m_{p}j_{n}m_{n}}\langle j_{n}m_{n}|(t_{-}\sigma_{\mu})^{+}|j_{p}m_{p}\rangle (-1)^{l_{n}+j_{n}-m_{n}} \hat{c}^{\dagger}_{j_{p}m_{p}} \hat{c}^{\dagger}_{j_{n}-m_{n}},
\end{equation}

was taken into account. The interaction constants $\chi$ and
$\kappa$ in units of MeV were both taken to be positive. The
different signs of $V^{pp}$ and $V^{ph}$ reflect a well-known
feature of the nucleon-nucleon interaction; namely, that the ph
force is repulsive while the pp force is attractive. Instead of
using a parametrization of chi and kappa values as a function of
nucleon number, we chose to fix specific values of chi and kappa for
each isotopic chain. Example giving in previous pn-QRPA calculation
Homma and collaborators  took $\chi = 5.2/A^{0.7}$ MeV and $\kappa =
0.58/A^{0.7}$ MeV \cite{Hom96}. These values were deduced from a fit
to experimental half-lives and for every isotopic chain fixed values
of chi and kappa allowed to deduce a locally best value of chi and
kappa (see also Ref. \cite{Hir93} which uses the same recipe). For
further study of effect of interaction constants, $\chi$ and
$\kappa$, on the pn-QRPA calculations, we refer to \cite{Hir93,
Sta90a}. It was later shown that fixing values of $\chi$ and
$\kappa$ for an isotopic chain led to better reproduction of
experimental data \cite{Nab12, Nab11}. For the case of nickel
 and iron isotopes the values of interaction constants were taken accordingly from Ref.
\cite{Nab12} and Ref. \cite{Nab11}, respectively.  The values of
$\chi $ and $\kappa $, along with the value of deformation
parameter, used in the current work, are shown in Table~\ref{ta1}.

Using a separable interaction allows  the pn-QRPA calculations
to be solved in a much larger single-particle basis than with a
general/semi-realistic interaction; in this case we used up to 7
$\hbar \omega$ shells. Such calculations have been used extensively
in computing GT transitions for astrophysical applications for a
wide variety of nuclide (e.g. \cite{Nab12, Nab05, Nab08, Nab09})

Matrix elements of the forces which appear in RPA
equation~(\ref{RPA_A}),(\ref{RPA_B}) are separable,
\begin{equation}
V^{ph}_{pn,p^{\prime}n^{\prime}} = +2\chi
f_{pn}(\mu)f_{p^{\prime}n^{\prime}}(\mu),
\end{equation}
\begin{equation}
V^{pp}_{pn,p^{\prime}n^{\prime}} = -2\kappa
f_{pn}(\mu)f_{p^{\prime}n^{\prime}}(\mu),
\end{equation}
with
\begin{equation}
f_{pn}(\mu) =
\sum_{j_{p}j_{n}}D^{m_{p}\alpha_{p}}_{j_{p}}D^{m_{n}\alpha_{n}}_{j_{n}}\langle j_{p}m_{p}|t_{-}\sigma_{\mu}|j_{n}m_{n} \rangle,
\end{equation}
which are single-particle GT transition amplitudes defined in the
Nilsson basis. For the case of separable forces, the matrix
equation~(\ref{RPA}) reduces to an algebraic equation of second
order (when $\kappa$ = 0) and with a finite value of $\kappa$ it
transforms to a fourth order equation. Methods of finding roots of
these equations can be seen in Ref. \cite{Mut92}. For details on
QRPA model parameters we refer to \cite{Nab12}.

The purpose of this paper is to compare general trends of these
calculations with the $0\hbar\omega$ calculations using a more
general, realistic interaction described in the previous section.

\section{Results and comparison}

Using the Gamow-Teller operator $\vec{\sigma} \tau_\pm$ yields
the Ikeda sum rule \cite{Ike64} for a parent nucleus with $Z$ protons
and $N$ neutrons:
\begin{equation}
\sum B(GT_-) - \sum B(GT_+) = 3(N-Z),
\end{equation}
in our calculations. For use in astrophysical reaction rates, and to
compare to experimental data, and to prior calculations, we multiply our
results by a quenching factor of 0.6  \cite{Vet89} typical for
nuclei.  Note that we do not include the axial weak coupling
constant $g_A$ as the published data and calculations we compare to also
leave it out, see for example Eq.~1 in \cite{Suz09}, which uses a 
quenching factor of $(0.74)^2=0.55$.

% we have to multiply our calculated
%strengths by $(g_A / g_V)^2 = (-1.254)^2$ \cite{Tow95}, and then by
%an additional quenching factor of 0.6 \cite{Vet89} typical for
%nuclei.

The ultimate goal is to provide reliable weak rates for
astrophysical environments, many of which cannot be measured
experimentally. Even theoretically this is a complex and difficult
issue. For example, $\beta$-decay and capture rates are
exponentially sensitive to the location of GT$_{+}$ resonance while
the total GT strength affect the stellar rates in a more or less
linear fashion \cite{Auf96}. In sufficiently hot astrophysical
environments one must include rates with an excited parent state.
But rates off excited states are difficult to get: an $(n,p)$
experiment on a nucleus $(Z,A)$ shows where in $(Z-1,A)$ the
GT$_{+}$ centroid corresponding only to the ground state of $(Z,A)$
resides. The calculations described in this paper are also limited
only to ground state parents, although we hope to tackle excited
parents in the future. For a discussion of calculation of excited
state GT strength functions we refer to \cite{Cau99} using the shell
model and \cite{Nab04} using the pn-QRPA model.

For this paper we focus on the variation in Gamow-Teller strengths from different
RPA calculations, to give us an idea of the theoretical uncertainty.

Table~\ref{ta3} shows the mutual comparison of the various RPA
models used in this project. This table shows the  values of the
centroids, widths and total strength values of the calculated and
measured (where available) GT distributions, both in $\beta^{-}$ and
electron capture directions, for various iron-regime nuclei.  It is
clear from Table~\ref{ta3} that the GXPF1 interaction calculates the
biggest value of the total GT strength. On the other end the 
pn-QRPA tends to calculate lower total strength values.
Regarding the calculation of centroids in various RPA models, we
note that the pn-QRPA calculated centroid resides at lower energy in
daughter, except for the case of $^{55}$Fe where the KB3G
interaction calculates the lowest centroids in the electron capture
direction. On the other extreme the GXPF1 calculates the highest
centroids except for the case of $^{55}$Fe where the pn-QRPA model
tops the chart in the electron capture direction. One also notes
that the pn-QRPA calculated GT strength distributions tend to have a
larger width.
%restored for last version
For related discussion on the pn-QRPA built on a
deformed self-consistent mean field basis obtained from two-body
density-dependent Skyrme forces for iron mass region we refer to
\cite{Sar03}.
The calculated GT strength distributions using
different interactions will next be discussed below.

An obvious question would be how the various RPA calculations compare
with the measured data as well as full shell-model diagonalization.
Thus we also show in Table~\ref{ta3} how the values of calculated GT
centroids and total GT strengths (both in $\beta^{-}$ and electron
capture directions), using various RPA models, compare with the
available experimental data. For the sake of comparison we also
include the shell model calculation using the GXPF1J Hamiltonian
\cite{Hon05} taken from Table~I of Ref. \cite{Suz09} (the centroid
value was not available).   The authors claimed that the GXPF1J
interaction leads to spreading of calculated strength and better
reproduction of observed strength in Fe and Ni isotopes (see also
Ref. \cite{Suz11}). Overall, the differences between the shell-model diagonalization
strengths and those from RPA calculations using shell-model interactions 
is similar to that previously reported \cite{stetcu2004}.
The experimental centroids and widths were calculated from
the reported measured data and all measured data are given to one
decimal place in Table~\ref{ta3}. For the $\beta^{-}$ side the
measured data for $^{54}$Fe were taken from Refs.
\cite{Rap83,Vet89,And90,Ada12}. It is to be noted that the recent
high-resolution ($^{3}$He, $t$) charge-exchange reaction on
$^{54}$Fe performed by Adachi and collaborators \cite{Ada12} report
a much lower value of $\sum B(GT_-)$ = 4.00 $\pm$ 0.37 up to 12 MeV
in $^{54}$Co. Ref. \cite{Rap83} reported  the total strength of
$\sum B(GT_-)$ = 7.8 $\pm$ 1.9 and was not able to calculate B(GT)
values at discrete excitation energies beyond 4.5 MeV in daughter.
Hence it was not possible for us to calculate the centroid and width
in this particular case. For the electron capture direction,
experimental data for $^{54}$Fe were taken from Refs.
\cite{Vet89,Roe93} as also mentioned in Table~\ref{ta3}. Measured
data for $^{56}$Fe in the electron capture direction were taken from
Refs. \cite{Roe93,Elk94} while for the $\beta^{-}$ direction we only
quote the reported value of $\sum B(GT_-)$ = 9.9 $\pm$ 2.4 by
Rapaport and collaborators \cite{Rap83}. The authors were unable to
extract GT strengths for discrete excited states beyond 5.9 MeV in
$^{56}$Co making it impossible for us to calculate the centroid and
width in this case.

Recently $(p,n)$ charge-exchange reaction in inverse kinematics at
intermediate energies were used to extract GT strengths for the
unstable nucleus $^{56}$Ni \cite{Sas11}. The authors reported a
value of $\sum B(GT_-)$ = 3.5 $\pm$ 0.3 and due an additional
uncertainty in GT unit cross section (normalization factor of B(GT))
also quoted a value of $\sum B(GT_-)$ = 3.8 $\pm$ 0.2 in Ref.
\cite{Sas12}. For the case of $^{58}$Ni, measured data were taken
from Refs. \cite{Elk94,Col06} for the electron capture direction.
Along the $\beta^{-}$ direction, authors in Ref. \cite{Rap83} quoted
the total strength of $\sum B(GT_-)$ = 7.5 $\pm$ 1.8 and were unable
to extract GT strengths for discrete excited states beyond 6.4 MeV
in $^{58}$Cu (accordingly we were unable to calculate centroid and
width for this case in Table~\ref{ta3}). Fujita and collaborators
extracted GT strength using the ($^{3}$He,t) reaction up to 8.3 MeV
in $^{58}$Cu \cite{Fuj02} and later up to higher daughter energies
of 13 MeV \cite{Fuj07}.

It is to be noted that we used a quenching factor of 0.6 for the
calculated GT strength using the pn-QRPA model \cite{Nab11} which is
normally done in stellar weak rate calculations and also discussed
in Section~1. Note that the shell model interactions \cite{Suz09}
calculated strengths were quenched by a universal quenching factor
of 0.55 rather than 0.6. Table~\ref{ta3} shows that the pn-QRPA
model calculates the centroid at a much lower energy than other
shell model interactions. Further in all cases it is seen that the
pn-QRPA model best reproduces the placement of measured centroid.
The comparison is exceptionally good for the case of GT$_{-}$
centroid of $^{54}$Fe and for the GT$_{+}$ centroids of $^{56}$Fe
and $^{58}$Ni. On the other hand the GXPF1 interaction calculates
the highest centroid in daughter nuclei. The shell model
interactions calculate much better total strengths in comparison
with measured values. One should also keep in mind the uncertainties
present in measurements where various energy cutoffs are used as a
reasonable upper limit on the energy at which GT strength could be
reliably related to measured $\Delta L =0$ cross-sections as well as
slightly different values of quenching factor used in pn-QRPA and
shell model interactions before comparing the calculated numbers
with experimental data. The Ikeda sum rule is
satisfied  as seen in previous RPA calculations \cite{stetcu2004}.

\begin{figure}[htbp]
\caption{Calculated GT strength distributions in the electron
capture direction for $^{54}$Fe using different pn-RPA scenarios.
For details of interactions used see text. In inset measured data of
Ref. \cite{Roe93} is shown. The abscissa represents energy in
daughter. } \label{fig1}
\begin{center}
\begin{tabular}{c}
\includegraphics[width=0.8\textwidth]{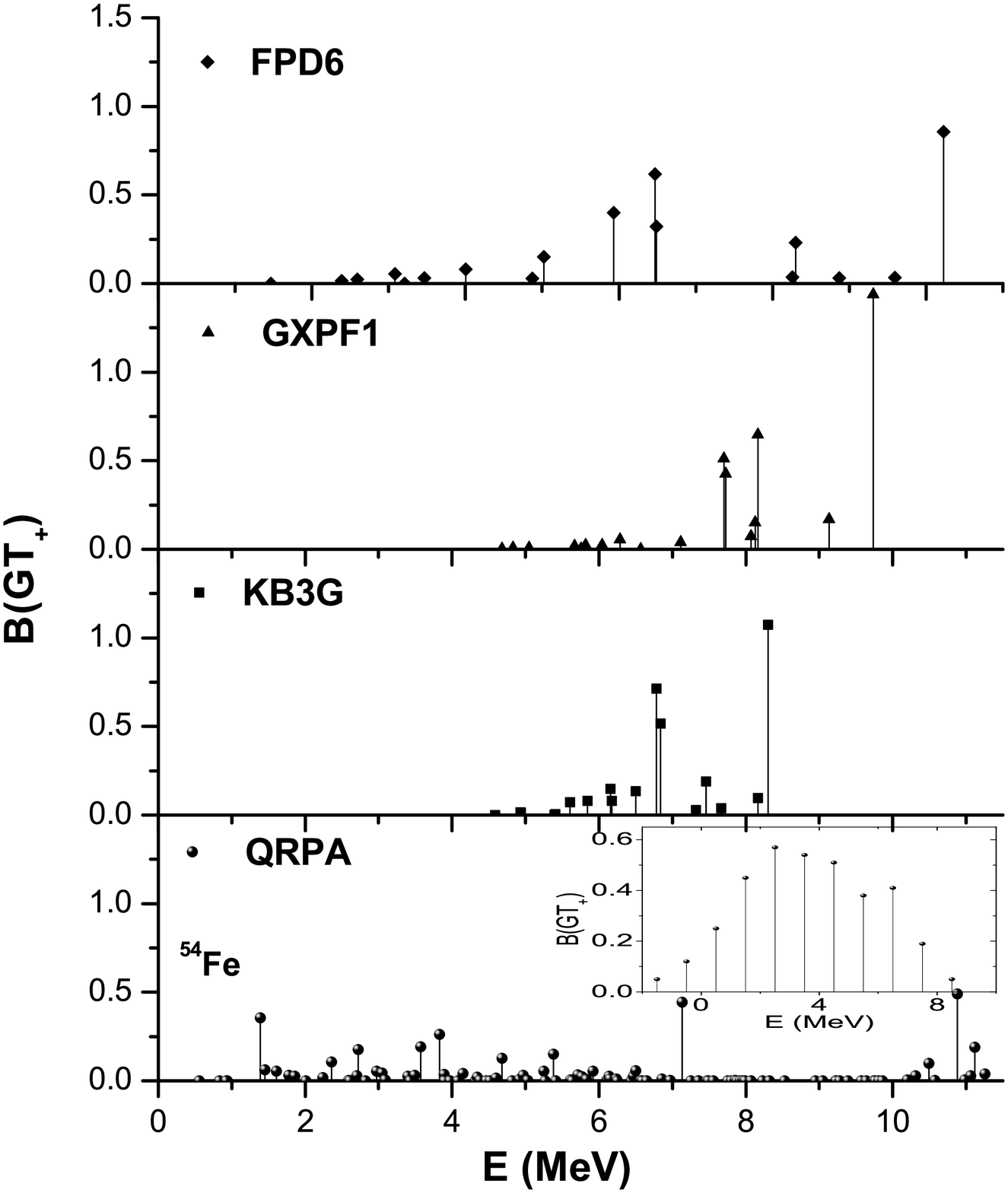}
\end{tabular}
\end{center}
\end{figure}

\begin{figure}[htbp]
\caption{Same as in Fig.~\ref{fig1} but for calculated GT strength
distributions in the $\beta^{-}$-decay direction. In inset the
recent measured data of Ref. \cite{Ada12} is shown.} \label{fig2}
\begin{center}
\begin{tabular}{c}
\includegraphics[width=0.8\textwidth]{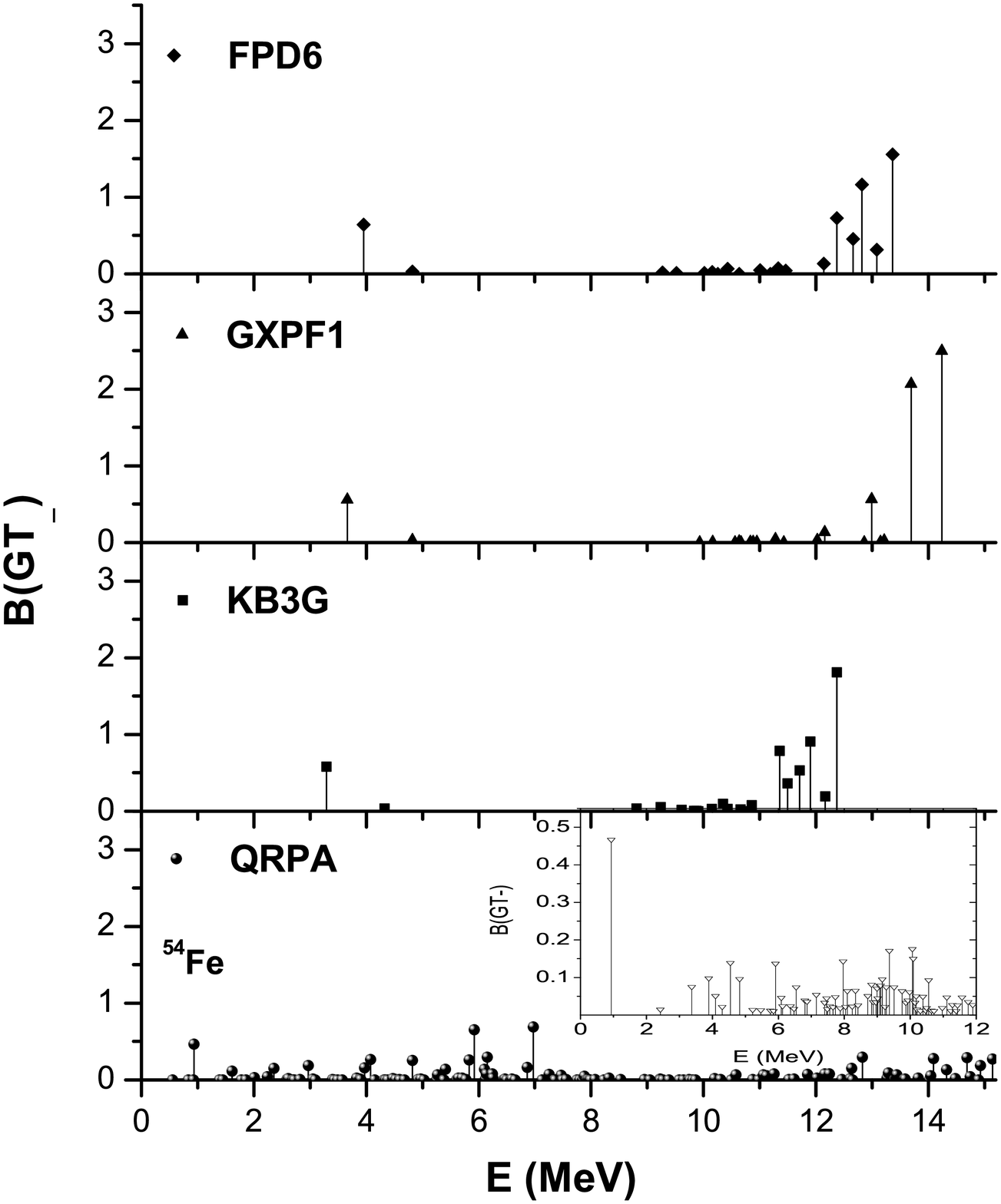}
\end{tabular}
\end{center}
\end{figure}

\begin{figure}[htbp]
\caption{Cumulative GT strength distributions for $^{54}$Fe using
different pn-RPA scenarios in the $\beta^{-}$-decay direction. For
details of interactions used see text. Experimental data (Exp) is
taken from Ref. \cite{Ada12}. The abscissa represents energy in
daughter.} \label{fig2a}
\begin{center}
\begin{tabular}{c}
\includegraphics[width=0.8\textwidth]{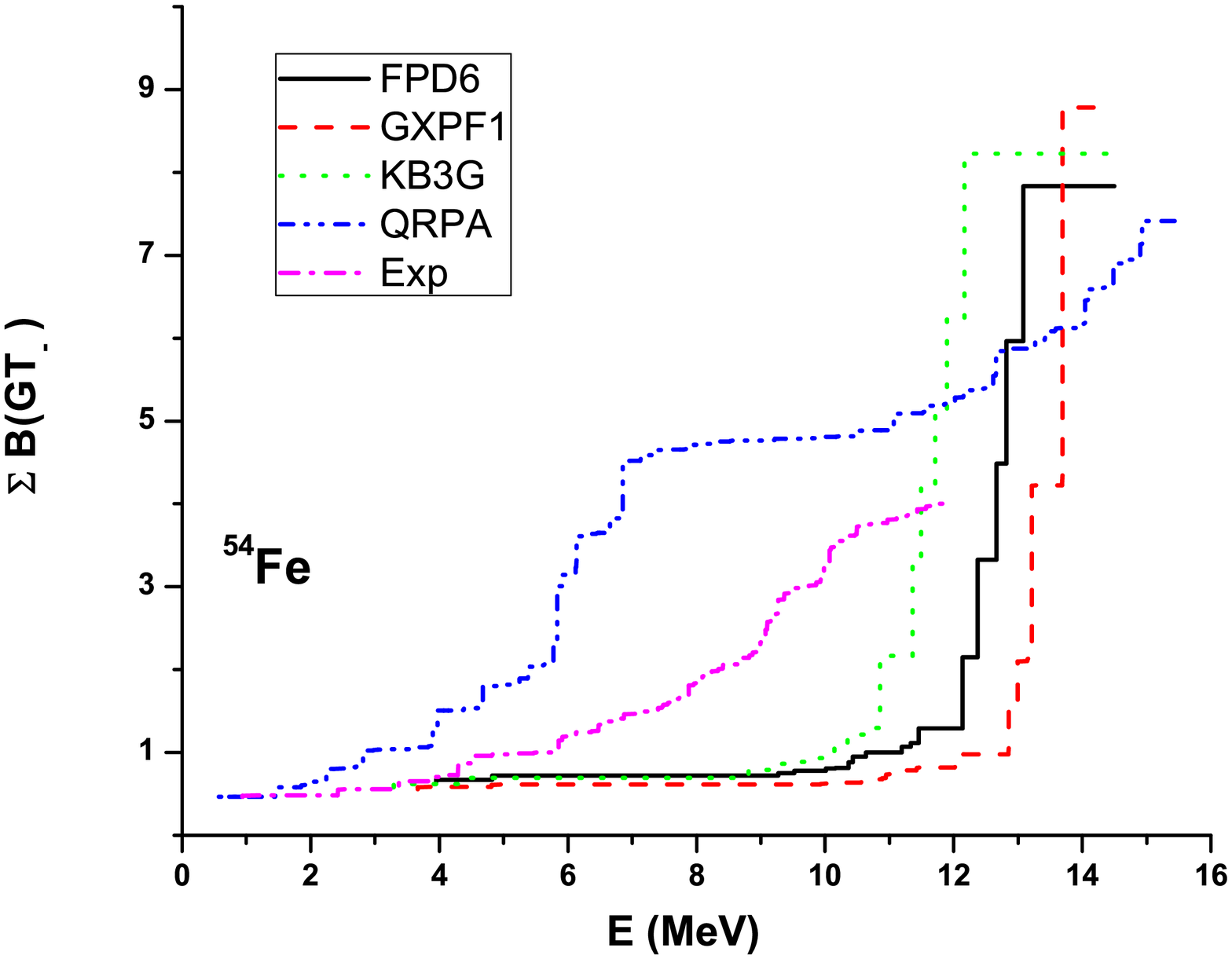}
\end{tabular}
\end{center}
\end{figure}

Fig.~\ref{fig1} and Fig.~\ref{fig2} show the calculated GT strength
distribution of $^{54}$Fe in various RPA calculations. Fig.~\ref{fig1}
displays the calculated GT strength distribution in the electron
capture direction whereas Fig.~\ref{fig2} shows similar calculation
in the $\beta^{-}$ direction. In the inset of Fig.~\ref{fig1} we
also show the $^{54}$Fe(n,p) reaction, measured at 97 MeV for
excitation energies in $^{54}$Mn \cite{Roe93}. Similarly we also
show the recently measured high-resolution ($^{3}$He, $t$) data on
$^{54}$Fe by Adachi and collaborators \cite{Ada12} in the inset of
Fig.~\ref{fig2}. Fig.~\ref{fig2a} shows the cumulative strength
distributions, in the $\beta^{-}$ direction, for various RPA
calculations and measured data \cite{Ada12}. Fig.~\ref{fig2a}
depicts the mutual comparison of various calculations with measured
data in a better fashion. It is noted that the pn-QRPA model
calculates GT transitions at low excitation energy. It is further
noted that the QRPA calculated distribution is better fragmented
than other RPA models and follows the trend of the measured data,
albeit with a much higher magnitude of strength distribution.

\begin{figure}[htbp]
\caption{Cumulative GT strength distributions for $^{55}$Fe using
different pn-RPA scenarios. For details of interactions used see
text. The abscissa represents energy in daughter.} \label{fig3}
\begin{center}
\begin{tabular}{c}
\includegraphics[width=0.8\textwidth]{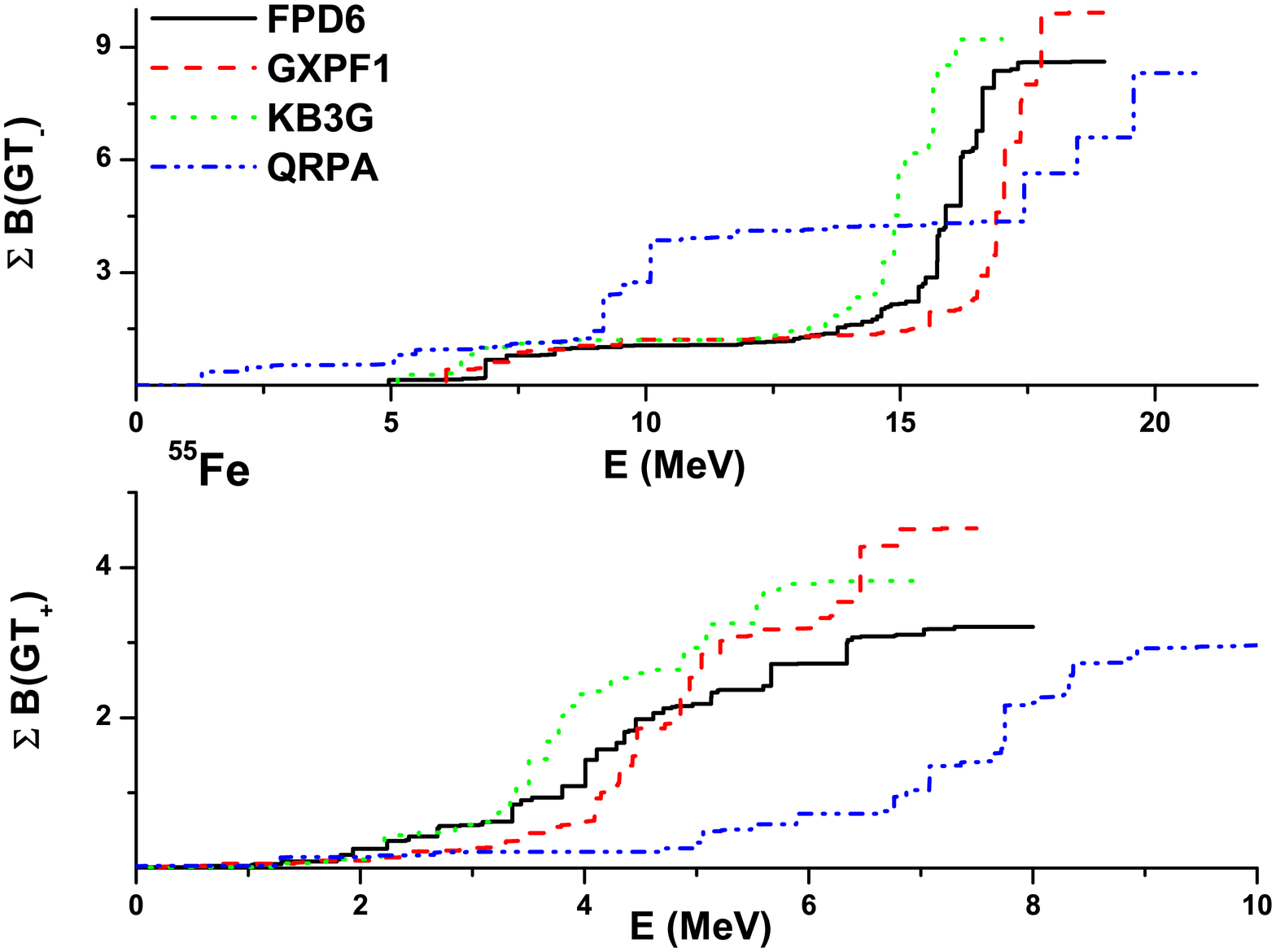}
\end{tabular}
\end{center}
\end{figure}

For the remaining cases we decided to show only the cumulative GT
strength distributions to save space. Fig.~\ref{fig3} display the
cumulative strength distributions in a two-panel frame. The upper
panel shows the cumulative strength distribution in the $\beta^{-}$
direction using the different RPA models whereas the lower panel
displays the results in the electron capture direction for the case
of $^{55}$Fe. The pn-QRPA calculates high-lying transitions in
$^{55}$Co (upper panel).  The KB3G interaction saturates first to
its maximum strength in both directions and gives the lowest values
for the centroid of the GT$_{+}$ distribution function. No measured
GT strength of $^{55}$Fe was available in literature to compare with
the different calculations.

\begin{figure}[htbp]
\caption{Cumulative GT strength distributions for $^{56}$Fe using
different pn-RPA scenarios. For details of interactions used see
text. Experimental data (Exp) is taken from Ref. \cite{Elk94}. The
abscissa represents energy in daughter.} \label{fig4}
\begin{center}
\begin{tabular}{c}
\includegraphics[width=0.8\textwidth]{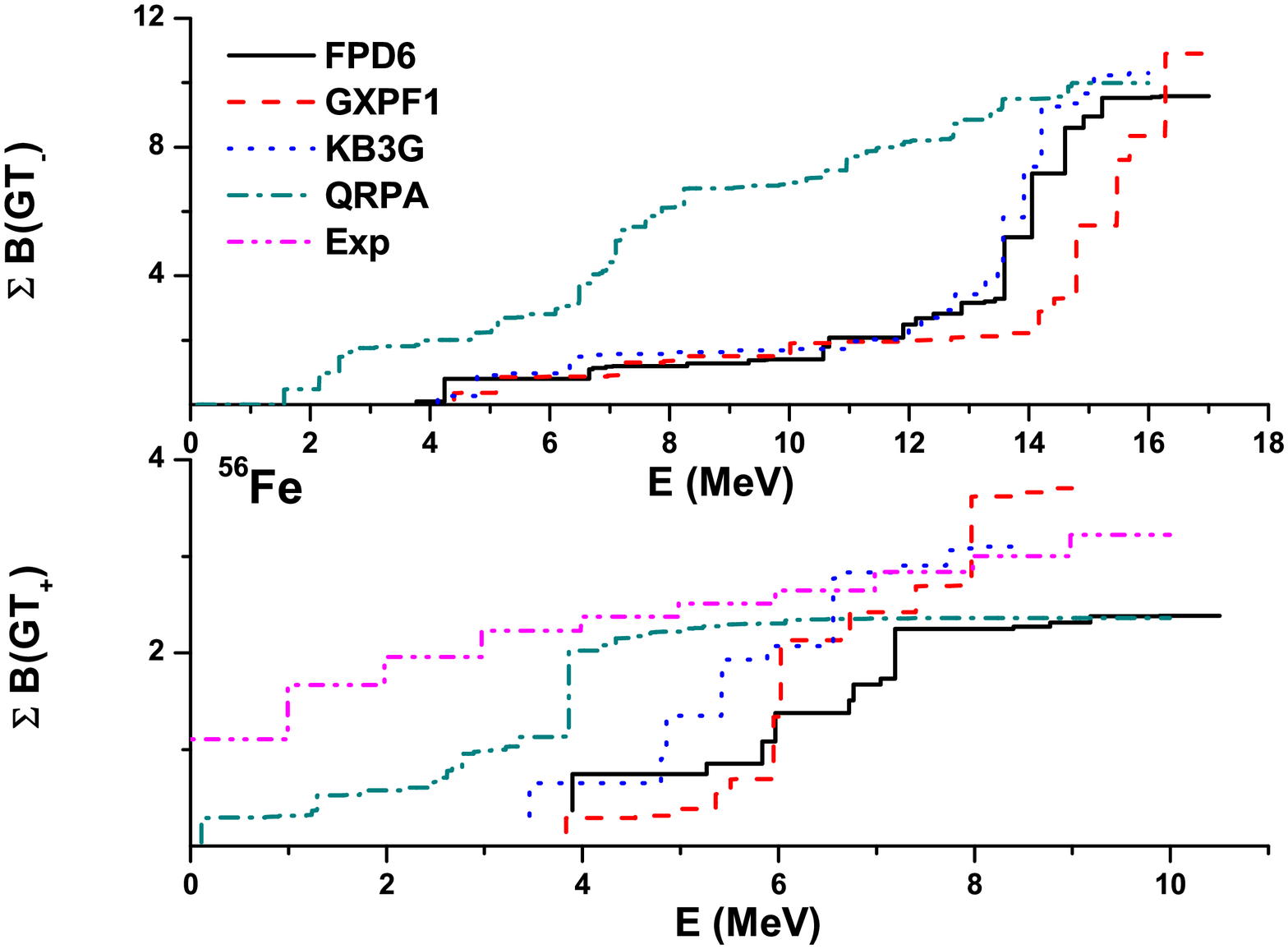}
\end{tabular}
\end{center}
\end{figure}

A similar comparison of cumulative strength functions for the case
of $^{56}$Fe is shown in Fig.~\ref{fig4}. Here we note that for the
$\beta^{-}$ direction the measured GT strength distribution quote
the value of $\sum B(GT_-)$ = 9.9 $\pm$ 2.4 up to 15 MeV in daughter
but extracts the B(GT) only up to discrete daughter excited states
of 3.5 MeV \cite{Rap83}. Accordingly it was not possible for us to
show the measured data in upper panel of Fig.~\ref{fig4}. In the
lower panel the measured data of Ref. \cite{Elk94} is shown for the
electron capture direction of $^{56}$Fe. Fig.~\ref{fig4} reveals
that the pn-QRPA model peaks at a faster pace compared to other RPA
models. Correspondingly the pn-QRPA calculates the lowest values for
the centroids in both directions for $^{56}$Fe (see
Table~\ref{ta3}). For the electron capture direction (lower panel)
the pn-QRPA best mimics the trend in the measured GT data. It can be
seen that the GT strength resides at much higher energies in the
daughter nuclei in the $\beta^{-}$ direction for all models.

\begin{figure}[htbp]
\caption{Cumulative GT strength distributions for the $N = Z$
nucleus $^{56}$Ni using different pn-RPA scenarios in the
$\beta^{-}$-decay direction. For details of interactions used see
text. Experimental data (Exp) is taken from Ref. \cite{Sas12}. The
abscissa represents energy in daughter.} \label{fig5}
\begin{center}
\begin{tabular}{c}
\includegraphics[width=0.8\textwidth]{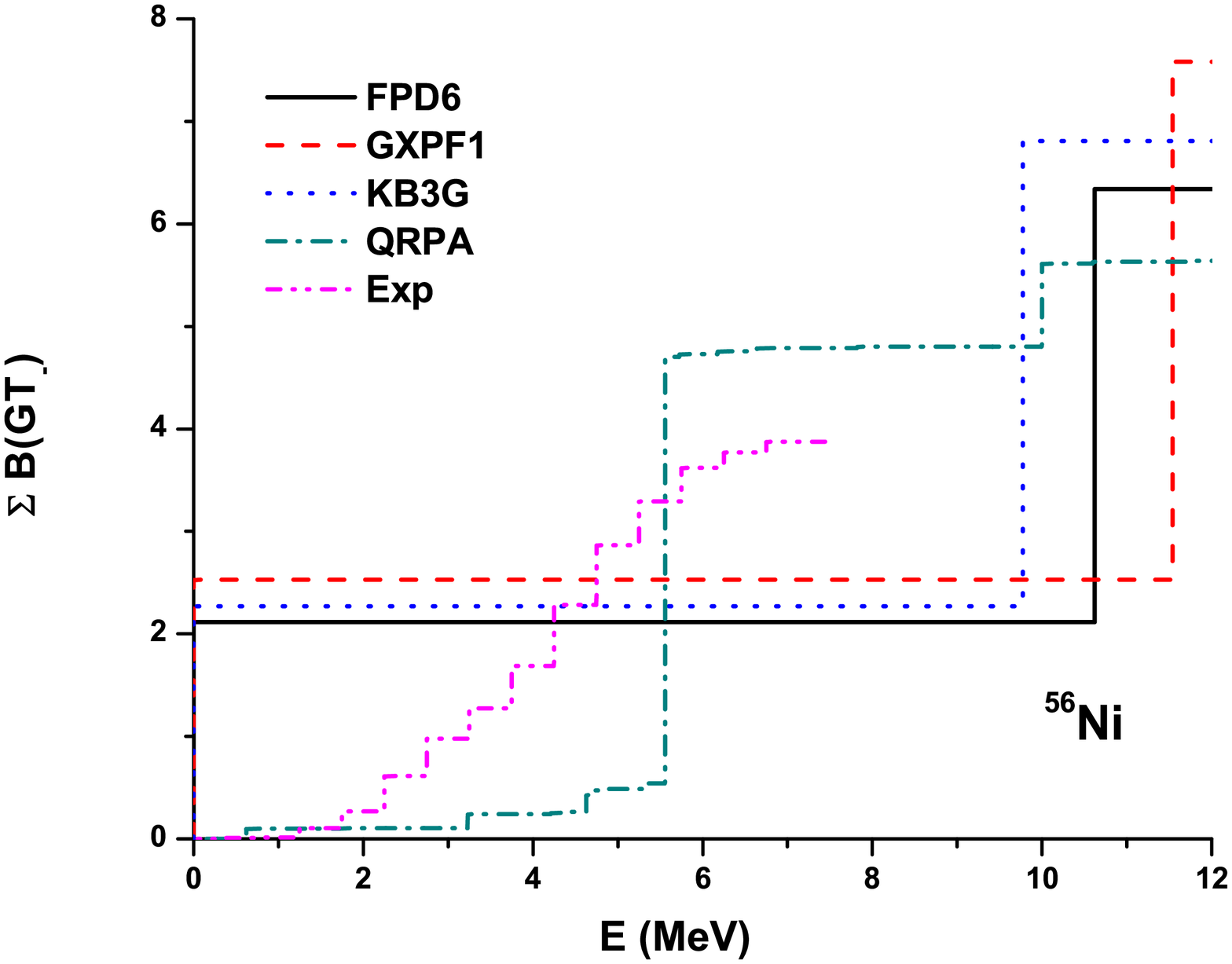}
\end{tabular}
\end{center}
\end{figure}

Fig.~\ref{fig5} shows the GT strength distribution of $N = Z$
nucleus $^{56}$Ni. Experimental data were taken from Ref.
\cite{Sas12}. Here one notes that for the models used, all strength
resides at one energy level in daughter nucleus for the FPD6, GXPF1
and KB3G interactions. This energy is 10.6, 11.5 and 9.8 MeV for the
FPD6, GXPF1 and KB3G interactions, respectively. The recent shell
model calculation by Suzuki et al. \cite{Suz09}, using the GXPF1J
interaction, showed that the GT strength distribution in $^{56}$Ni
is more fragmented and different from previous shell model
calculations resulting in enhanced production yields of heavy
elements. The authors further commented that the calculated GT
strength using the GXPF1J interaction was found to be more
fragmented with a remaining tail in the high excitation energy
compared with that obtained by the KB3G interaction. For the case of
$^{56}$Ni their calculated strength was fragmented into two peaks
whereas the total calculated strength was 11.32 (unquenched). On the
other hand we note from Fig.~\ref{fig5} that the pn-QRPA calculated
strength is still more fragmented over a range of energies with two
distinct peaks at 5.7 MeV and 10.6 MeV in the daughter nucleus. The
consequences of calculating a much more fragmented strength
distribution for $^{56}$Ni, using the pn-QRPA model, may also have
interesting scenario for heavy element nucleosynthesis and requires
further attention in nuclear network calculations.

\begin{figure}[htbp]
\caption{Cumulative GT strength distributions for $^{58}$Ni using
different pn-RPA scenarios. For details of interactions used see
text. Experimental data (Exp) is taken from Ref. \cite{Fuj02,Fuj07}
for the $\beta^{-}$-decay direction and from Ref. \cite{Col06} for
the electron capture direction. The abscissa represents energy in
daughter.} \label{fig6}
\begin{center}
\begin{tabular}{c}
\includegraphics[width=0.8\textwidth]{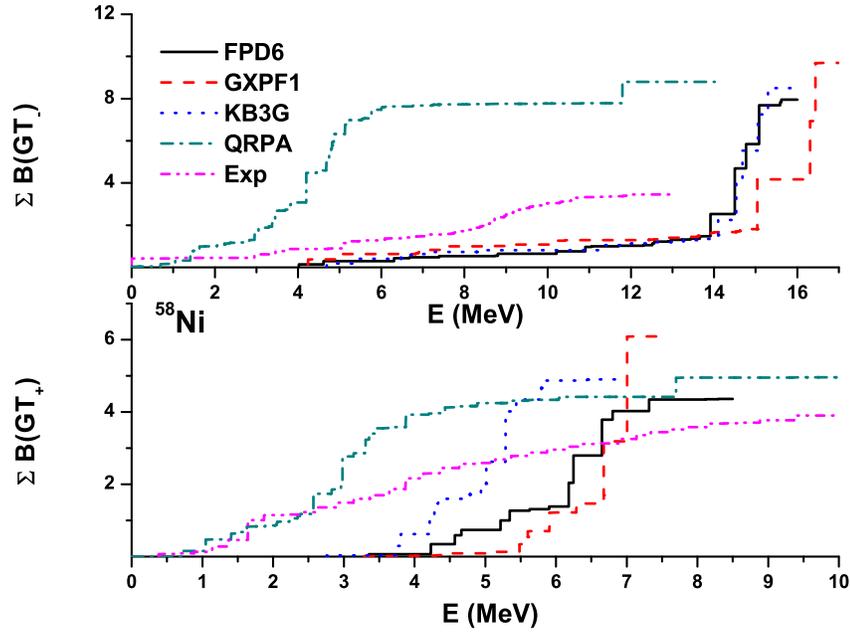}
\end{tabular}
\end{center}
\end{figure}

The cumulative strength distributions for the case of $^{58}$Ni is
shown finally in Fig.~\ref{fig6}. The upper panel shows the data for
the $\beta^{-}$ direction of $^{58}$Ni. Measured data were taken
from the charge-exchange ($^{3}$He,t) reaction performed by Fujita
and collaborators \cite{Fuj02,Fuj07} with a total strength of $\sum
B(GT_-)$ = 3.5. A much higher total strength of $\sum B(GT_-)$ = 7.5
$\pm$ 1.8 was measured by Rapaport and collaborators \cite{Rap83}.
However their measured data could not be presented in
Fig.~\ref{fig6} for reasons already mentioned before. The upper
panel also shows that the FPD6, GXPF1 and KB3G models calculate
strength up to higher energies in $^{58}$Cu as compared to the
pn-QRPA model. Consequently the pn-QRPA models calculate the
GT$_{-}$ centroid at a much lower energy of around 5 MeV in
$^{58}$Cu which also compares well with the experimental value of
6.9 MeV. The GXPF1 calculated centroid is thrice the value
calculated by the pn-QRPA model and can bear strong astrophysical
consequences. The lower panel depicts the data for the electron
capture direction of $^{58}$Ni. Here we also show the  measured data
by Cole and collaborators \cite{Col06} who extracted a total
strength of $\sum B(GT_+)$ = 4.1 $\pm$ 0.3 (it is to be noted that
the previous measured data by El-Kateb and co-workers  also measured
a total strength of 4.0 \cite{Elk94}) Once again the pn-QRPA model
calculates the lowest whereas the GXPF1 model calculates the highest
values for the GT$_{+}$ centroid. The pn-QRPA calculated placement
of centroid (3.6 MeV) compares well with the measured centroid of
4.4 MeV. It is the case of $^{58}$Ni where one sees the largest
difference in the placement of centroid using the QRPA and various
RPA interactions.

\section{Summary and conclusions}
Under astrophysical conditions, both the electron capture and beta
decay of fp-shell nuclei depend heavily on the centroid placement
and total strength of the calculated Gamow-Teller strength
distributions. In this work we  presented a comparative study of the
Gamow-Teller strength distributions  for a suite of astrophyiscally
important fp-shell nuclide ($^{54,55,56}$Fe,  and $^{56,58}$Ni)
using a suite of interactions, including semi-realistic interactions
in the $1p$-$0f$ space with the RPA and a separable multi-shell
interaction in the quasi-particle RPA. Where possible, we also
presented comparison with measured data. We further compared and
contrasted the statistics of calculated and measured GT strength
functions using various pn-RPA schemes in this paper. Our
calculations satisfied the model independent Ikeda sum rule. Work is
currently in progress for other important odd-A and odd-odd cases.

The QRPA model places the centroid at much lower energies in
daughter nuclei as compared to other RPA interactions. Further the
placement of GT centroids by the pn-QRPA model is, in general, in
good agreement with the centroids of the measured data. This
tendency of QRPA model can favor higher values of electron capture
rates in stellar environment and can bear significance for
astrophysical problems. On the other extreme, the GXPF1 interaction
usually leads to placement of GT centroid at much higher energies in
daughter compared to other pn-RPA interactions.

The present study showed that the total strengths, using various RPA
interactions, were in better agreement with the measured data when
compared to the QRPA calculated strength. Further the width of the
strength functions calculated within the QRPA was much
larger than those calculated with other RPA interactions. For the
special $N = Z$ nucleus $^{56}$Ni, the QRPA model calculated
Gamow-Teller strength function was well fragmented as compared to
other RPA interactions (including the recently used GXPF1J
interaction)  and may lead to interesting consequences for heavy
element nucleosynthesis.

\vspace{0.5in} \textbf{Acknowledgments}

JUN acknowledges the
support of research grant provided by the Higher Education
Commission, Pakistan, through HEC Project No. 20-1283.
CWJ was supported by the U.S.~Department of Energy for this
investigation through grant DE-FG02-96ER40985.

\newpage
%Tables (total 2)
\begin{table}
\caption{Values of GT strength force constants and deformation
parameters used in the present pn-QRPA calculation} \label{ta1}
\begin{center}
\begin{tabular}{cccc} \hline\\
Nucleus & $\chi$ & $\kappa$ & $\delta$ \\ \hline
%Cr & 0.2000 & 0.0000 & \\
%Mn & 0.5000 & 0.0850 & \\
$^{54}$Fe  & 0.15 & 0.07 & 0.195\\
$^{55}$Fe  & 0.15 & 0.07 & 0.083\\
$^{56}$Fe  & 0.15 & 0.07 & 0.239\\
%Co & 0.2000 & 0.0070 & \\
$^{56}$Ni & 0.001 & 0.052 & 0.011\\
$^{58}$Ni & 0.001 & 0.052 & 0.183\\\hline
\end{tabular}
\end{center}
\end{table}

\begin{table}
\caption{Statistics of calculated GT strength distributions for
nuclei using different RPA models given in second column. The GT
strength values for shell-model calculations using GXPF1J
interaction were taken from Table~I of Ref. \cite{Suz09}, while
experimental references are: (a) $\rightarrow$ \cite{Roe93}, (b)
$\rightarrow$ \cite{Vet89}, (c) $\rightarrow$ \cite{Ada12}, (d)
$\rightarrow$ \cite{Rap83}, (e) $\rightarrow$ \cite{And90}, (f)
$\rightarrow$ \cite{Elk94}, (g) $\rightarrow$ \cite{Sas12}, (h)
$\rightarrow$ \cite{Col06}, (i) $\rightarrow$ \cite{Fuj02,Fuj07}.
 } \label{ta3}
\begin{center}
\scriptsize\begin{tabular}{c|c|ccc|ccc} Nucleus &
Model & E(GT$_{+}$) & Width(GT$_{+}$) & $\sum B(GT_+)$  & E(GT$_{-}$) & Width(GT$_{-}$) &  $\sum B(GT_-)$ \\
& & MeV & MeV & arb. units & MeV & MeV & arb. units \\
\hline $^{54}$Fe & & & & & & &\\
& FPD6  & 7.50 & 1.01 & 4.23 & 11.92 & 2.56 & 7.83 \\
& GXPF1 & 8.42 & 1.00 & 5.19 & 12.97 & 2.59 & 8.79 \\
& KB3G  & 7.13 & 0.83 & 4.62 & 11.02 & 2.34 & 8.24 \\
& QRPA  & 5.80 & 3.38 & 3.59 & 8.03  & 4.40 & 7.41 \\
 & GXPF1J (SM) & -  & -  & 4.0 &  - & - & 7.3 \\
& EXP$^{(a)}$  &  3.7  & 2.2 & 3.5  & - & - & - \\
& EXP$^{(b)}$  &  3.0  & 2.3 & 3.1  & 8.3 & 3.6 & 7.5 \\
& EXP$^{(c)}$  &  -  & - & -  & 7.4 & 3.1 & 4.0 \\
& EXP$^{(d)}$  &  -  & - & -  & - & - & 7.8 \\
& EXP$^{(e)}$  &  -  & - & -  & 7.6 & 3.2 & 6.0 \\
\hline $^{55}$Fe & & & & & & &\\
& FPD6  & 4.42 & 1.43 & 3.21 & 14.87 & 3.02 & 8.61 \\
& GXPF1 & 5.07 & 1.26 & 4.52 & 15.93 & 3.24 & 9.92 \\
& KB3G  & 4.03 & 1.08 & 3.82 & 13.96 & 2.95 & 9.22 \\
& QRPA  & 7.12 & 1.75 & 2.98 & 13.68 & 5.77 & 8.31 \\
\hline $^{56}$Fe & & & & & & &\\
& FPD6  & 6.12 & 1.75 & 2.38 & 12.79 & 3.00 & 9.58 \\
& GXPF1 & 6.73 & 1.38 & 3.70 & 14.09 & 3.39 & 10.90 \\
& KB3G  & 5.49 & 1.32 & 3.10 & 12.54 & 3.13 & 10.30 \\
& QRPA  & 3.14 & 1.53 & 2.36 &  7.79 & 3.79 & 9.99 \\
 & GXPF1J (SM) & -  & -  & 2.9 &  - & - & 9.5 \\
& EXP$^{(a)}$  &  2.7  & 2.0 & 2.3  & - & - & - \\
& EXP$^{(d)}$  &  -  & - & -  & - & - & 9.9 \\
& EXP$^{(f)}$  &  3.5  & 3.2 & 3.2  & - & - & - \\
\hline $^{56}$Ni & & & & & & &\\
& FPD6  & 10.62 & 0.00 & 6.34 & 10.62 & 0.00 & 6.34 \\
& GXPF1 & 11.54 & 0.00 & 7.58 & 11.54 & 0.00 & 7.58 \\
& KB3G  & 9.77  & 0.00 & 6.81 & 9.77  & 0.00 & 6.81 \\
& QRPA  & 6.32 &  1.91 &  5.64 & 6.32  & 1.91 &  5.64 \\
& EXP$^{(g)}$  &  -  & - & -  & 4.1 & 1.4 & 3.8 \\
\hline $^{58}$Ni & & & & & & &\\
& FPD6  & 6.15 & 0.92 & 4.36  & 13.94 & 2.49 & 7.96 \\
& GXPF1 & 6.79 & 0.58 & 6.09 & 14.99 & 3.34 & 9.70 \\
& KB3G  & 5.02 & 0.70 & 4.91  & 13.97 & 2.55 & 8.51 \\
& QRPA  & 3.57 & 1.91 & 4.97  &  4.97 & 2.82 & 8.79 \\
 & GXPF1J(SM) & -  & -  & 4.7 &  - & - & 8.0 \\
& EXP$^{(d)}$  &  -  & - & -  & - & - & 7.5 \\
& EXP$^{(f)}$  &  4.0  & 2.1 & 4.0  & - & - & - \\
& EXP$^{(h)}$  &  4.4  & 5.0 & 4.1  & - & - & - \\
& EXP$^{(i)}$  &  -  & - & -  & 6.9 & 3.2 & 3.5 \\
\hline
\end{tabular}
\end{center}
\end{table}

\newpage
%figures (total 7)

\end{document}